\documentclass{aa}  

\usepackage{graphicx}
\usepackage{txfonts}
\usepackage[colorlinks=true,
            linkcolor=blue,
            urlcolor=blue,
            citecolor=blue]{hyperref}
\newcommand{\hmsun}{h^{-1}{\rm M}_\odot}
\newcommand{\hmpc}{h^{-1}{\rm Mpc}}

\begin{document} 
  
   \title{The evolution of the halo occupation distribution in cosmic voids.}

   \author{Ignacio G. Alfaro \thanks{E-mail:german.alfaro@unc.edu.ar}, Facundo Rodriguez \& Andr\'es N. Ruiz}
   
   \authorrunning{I. G. Alfaro et al.}
   
   \institute{Instituto de Astronomía Teórica y Experimental, CONICET-UNC, Laprida 854, X5000BGR, C\'ordoba, Argentina \\ Observatorio Astron\'omico de C\'ordoba, UNC, Laprida 854, X5000BGR, C\'ordoba, Argentina.}

   \date{\today}

  \abstract
   {The halo occupation distribution (HOD) is a cornerstone in understanding galaxy formation within the large-scale structure.
   On the other hand, in the context of the large-scale structure of the Universe, voids are regions with singular characteristics, given their extension, the low number of objects that inhabit their interior and their own dynamics.
   Furthermore, the HOD of these regions exhibits significant deviations from the average, as shown by several studies in semi-analytical models, hydrodynamic simulations, and observations. 
   }
   {This paper investigates the temporal evolution of the HOD in cosmic voids from redshift $z=0.5$ to the present day. We aim to understand how the void environment shapes galaxy occupation over time and to identify the factors driving the unique HOD observed in these underdense regions.}
   {We use the MDPL2-SAG semi-analytic galaxy catalog and identify spherical cosmic voids at different redshifts. We analyze the HOD within voids across ten simulation snapshots, comparing it with the global HOD. Furthermore, we trace the evolution of $z=0$ void galaxies and their host halos back in time, examining their formation times and local density evolution.}
   {Our results reveal that the HOD within voids is consistently lower than the average HOD at all redshifts considered. Tracking $z=0$ void galaxies, we find that their HOD remains lower than average throughout the redshift range, with no significant redshift evolution in this relative difference.  Analysis of halo formation histories shows that lower-mass void halos are younger, while massive void halos are older compared to the average. Spatially, lower-mass halos are found in the inner void regions, whereas massive halos are located closer to void boundaries. The local density of massive void halos decreases significantly towards $z=0$, contrasting with the relatively stable local density of lower-mass void halos.}
  {}

   \maketitle
  
%

\section{Introduction}
\label{sec:introduction}

The connection between galaxies and the dark matter halos they inhabit is a cornerstone of our understanding of galaxy formation and evolution within the large-scale structure of the Universe \citep{White1978}. 
The halo occupation distribution (HOD), which statistically describes the number of galaxies residing in halos of a given mass, has become a powerful tool for exploring this connection \citep[e.g.][]{Jing1998, Ma2000, Peacock2000, Seljak2000, Scoccimarro2001, Berlind2002, Cooray2002, Berlind2003, Zheng2005, Yang2007, Rodriguez2015}. 
Although the halo mass is the primary factor in determining their characteristics and galaxy population, recent studies show that the properties of the large-scale environment in which these halos reside also exert an influence \citep[e.g.][]{shi2018,Montero-Dorta_2024, shi2018}.
In particular, initial HOD models often assumed a primary dependence on halo mass, recent studies have revealed a significant influence of the large-scale environment on the way galaxies populate dark matter halos \citep[][among others]{Zehavi2018, Artale2018, Alfaro2020, Alfaro2021, alfaro_hod_2022, Perez2024}.
This environmental influence becomes more pronounced when focusing on regions with matter densities that significantly deviate from the mean.
Specifically, marked deviations in the HOD are found in extreme environments such as the densest regions and cosmic voids \citep{Alfaro2020, Alfaro2021, Perez2024}. 
Cosmic voids are characterized by their extremely low densities and reduced gravitational interactions and provide unique environments for studying galaxy evolution \citep{Rojas2004, Hoyle2005, Hoyle2012, Patiri2006, Ceccarelli2008}.
Previous studies have demonstrated that galaxies in voids are typically bluer, fainter, and have younger stellar populations with intense star formation activity \citep{Rojas2004, Rojas2005, Hoyle2005, Hoyle2012, jian2022, Rodriguez-Medrano2023}.
Focusing on the study of HOD in voids, \cite{Alfaro2020} found that halos within these regions are systematically populated by fewer galaxies.
This difference becomes more pronounced for halos of higher mass.
These results were consistent across two different types of simulation: a semi-analytic model and a hydrodynamical simulation.
Building on this work, \cite{alfaro_hod_2022} conducted a similar analysis using observational data from the Sloan Digital Sky Server - Data Release 12 \citep[SDSS-DR12;][]{Alam2015} spectroscopic catalog and the galaxy group catalog of \cite{RodriguezMerchan_2020}.
Despite the inherent observational biases, the results supported the predictions from simulations, demonstrating the influence of these vast underdense regions on how galaxies populate dark matter halos.
The aforementioned studies provided evidence for the impact of voids on the evolutionary history of halos, but these differences were analyzed in redshift $z=0$.
A natural question that arises from this context is: How does the HOD in these environments evolve over time?
Using the semi-analytic galaxy catalog MDPL2-SAG \citep{knebe_multidark_2018}, this paper aims to explore the environmental dependence of the HOD in greater depth by focusing on its temporal evolution within cosmic voids.
By studying the HOD at different epochs, we aim to understand the underlying mechanisms driving this process.
\textcolor{black}{We constrained our analysis between $z=0$ and $z=0.5$ following two critical factors. First, we did not expect a remarkable evolution in the void size function of our different redshift samples, ensuring consistent regions identification. Second, observational limitations increasingly complicate void and galaxy group detection at higher redshifts because of declining tracer density and increasing selection effects. These constraints motivate our focus on the local Universe, providing a robust baseline for understanding void galaxy populations.}
This paper is organized as follows: 
In Section \ref{sec:data}, we describe the data used in this work, including the MDPL2-SAG semi-analytic galaxy catalog and our void identification methodology.
Section \ref{sec:HOD_results} presents our analysis of how cosmic voids affect the HOD across different redshifts, examining both the global impact of voids and their specific effects on $z=0$ void galaxies. 
In Section \ref{sec:mass_analysis}, we analyze the formation and environmental evolution of void galaxies at $z=0$.
We examine the formation times of their host halos, their spatial distribution within voids, and the evolution of their local density.
These results provide insight into how the void environment influences galaxy formation and help explain the distinct HOD trends observed in these regions.
Finally, in Section \ref{sec:conclusions}, we summarize our main findings and present our conclusions.
%

\begin{table*}
\begin{center}
\begin{tabular}{ccccccc}
\hline\hline
Snapshot & Redshift & Total Galaxies & Total Halos & Total Voids & Total void galaxies & Total void halos \\ 
& & ($\times 10^6$) & ($\times 10^6$) & & ($\times 10^6$) & ($\times 10^6$) \\
\hline
125 & 0.000 & 39.04 & 32.83 & 20748 & 2.29 & 2.19 \\
124 & 0.022 & 38.76 & 32.45 & 20745 & 2.14 & 2.04 \\
123 & 0.045 & 38.39 & 32.04 & 22212 & 1.91 & 1.83 \\
122 & 0.069 & 37.98 & 31.63 & 21858 & 1.85 & 1.76 \\
121 & 0.093 & 37.57 & 31.24 & 21397 & 1.79 & 1.71 \\
120 & 0.117 & 37.19 & 30.88 & 23881 & 2.23 & 2.13 \\
117 & 0.194 & 35.97 & 29.75 & 26071 & 2.23 & 2.10 \\
113 & 0.304 & 34.36 & 28.30 & 27875 & 2.14 & 2.04 \\
110 & 0.394 & 33.11 & 27.20 & 29553 & 2.12 & 2.01 \\
107 & 0.490 & 31.81 & 26.09 & 29629 & 1.77 & 1.68 \\
\hline
\end{tabular}
\caption{Properties of the selected snapshots used in this work.Column 1: snapshot number from the MDPL2-SAG simulation. Column 2: corresponding redshift. Column 3: total number of galaxies that satisfy our selection criteria. Column 4: total number of halos that satisfy our selection criteria. Column 5: number of cosmic voids identified in each snapshot. Column 6: total number of galaxies identified within cosmic voids. Column 7: total number of halos identified within cosmic voids.}
\label{tab:snapshots}
\end{center}
\end{table*}

\section{Data}
\label{sec:data}

In this section, we describe the datasets used in this work to study the temporal evolution of the HOD in cosmic voids.
For this, we employ a catalog of semi-analytic galaxies and their corresponding dark matter halos, derived from the MultiDark Planck 2 (MDPL2) simulation and the Semi-Analytic Galaxies (SAG) model.
%

\subsection{MDPL2-SAG semi-analytic galaxy catalog}
\label{sec:data_mdpl}

The galaxy catalog used in this work is the MDPL2-SAG catalog \citep{knebe_multidark_2018}, publicly available in the database \textsc{CosmoSim}\footnote{\url{https://www.cosmosim.org}} \citep{riebe_multidark_2013}. The catalog is built up using the MDPL2 cosmological simulation \citep{prada_mdpl_2012,klypin_multidark_2016}, which follows the evolution of $3840^3$ dark matter particles
in a cubic box of $1000 \hmpc$ on side, with cosmological parameters consistent with Planck observations \citep{planck_2016}. 
Dark matter halos and subhalos were identified using the \textsc{Rockstar} algorithm \citep{behroozi_rockstar_2013}, and their merger trees were
constructed with \textsc{ConsistenTrees} \citep{behroozi_trees_2013}.
Dark matter halos were populated with galaxies using the semi-analytic model of galaxy formation and evolution \textsc{SAG} \citep{cora_sag_2018}.
This model includes the main physical processes relevant to galaxy formation, such as radiative cooling of gas, star formation, supernova feedback, 
chemical enrichment, growth of supermassive black holes, AGN feedback, and galaxy mergers.
The resulting catalog, known as MDPL2-SAG, has also been used in our previous studies of galaxy formation and evolution \citep{Alfaro2020, Alfaro2021}.
To study the evolution of galaxies and halos over time, we select 10 snapshots from the simulation, ranging from snapshot 125 to 107, 
which covers a redshift range from $z=0.0$ to $z=0.49$. The details of these snapshots, including the total number of galaxies and the identified
voids in each, are presented in Table \ref{tab:snapshots}
To ensure the resolution of the objects, we selected in each snapshot only those galaxies that reside in halos with masses $M_{200c}$ greater than $5 \times 10^{10} \hmsun$, where $M_{200c}$ is the mass contained in a sphere enclosing $200$ times the critical density of the Universe, $\rho_c = 2.7755 \times 10^{11} h^2 M_\odot{\rm Mpc^{-3}}$.
\textcolor{black}{The MDPL2 simulation uses dark matter particles with a mass of $8.271\times 10^{8} \hmsun$, and the halos we analyze contain a minimum of 50 particles, ensuring reliable mass estimates and structural properties.}
In addition, we only considered galaxies with stellar mass $M_\star$ greater than $5 \times 10^8 \hmsun$ and with absolute magnitudes in the r-band brighter than $M_r - 5 \log_{10}(h) = -16$. 
This selection guarantees a sample of well-resolved galaxies with reliable information along all the range of redshift selected.
It is important to note that in our analysis we exclude type-2 galaxies, also known as orphan galaxies.
These galaxies are those that have lost their subhalo and whose tracking in the simulation requires special treatment. 
\textcolor{black}{In the MDPL2-SAG catalog, the galaxy type classification follows a standard hierarchical criterion: type-0 (central galaxies) are defined as the most massive galaxy within a given dark matter halo, residing at the halo's center of mass. Type-1 (satellite galaxies) are all other galaxies within the same halo that are less massive than the central galaxy. This ensures that each halo contains exactly one central galaxy, with any additional galaxies classified as satellites.}
Since our main focus is the study of the HOD, we restrict ourselves to type-0 (central galaxies) and type-1 (satellite galaxies). 
Information on the type of galaxy was obtained directly from the MDPL2-SAG catalog.
With all these considerations, our final samples are presented in detail in Table \ref{tab:snapshots}.
%

\subsection{The cosmic voids catalogs}
\label{sec:data_voids}

\begin{figure}[ht!]
\begin{center}
\includegraphics[width=\columnwidth]{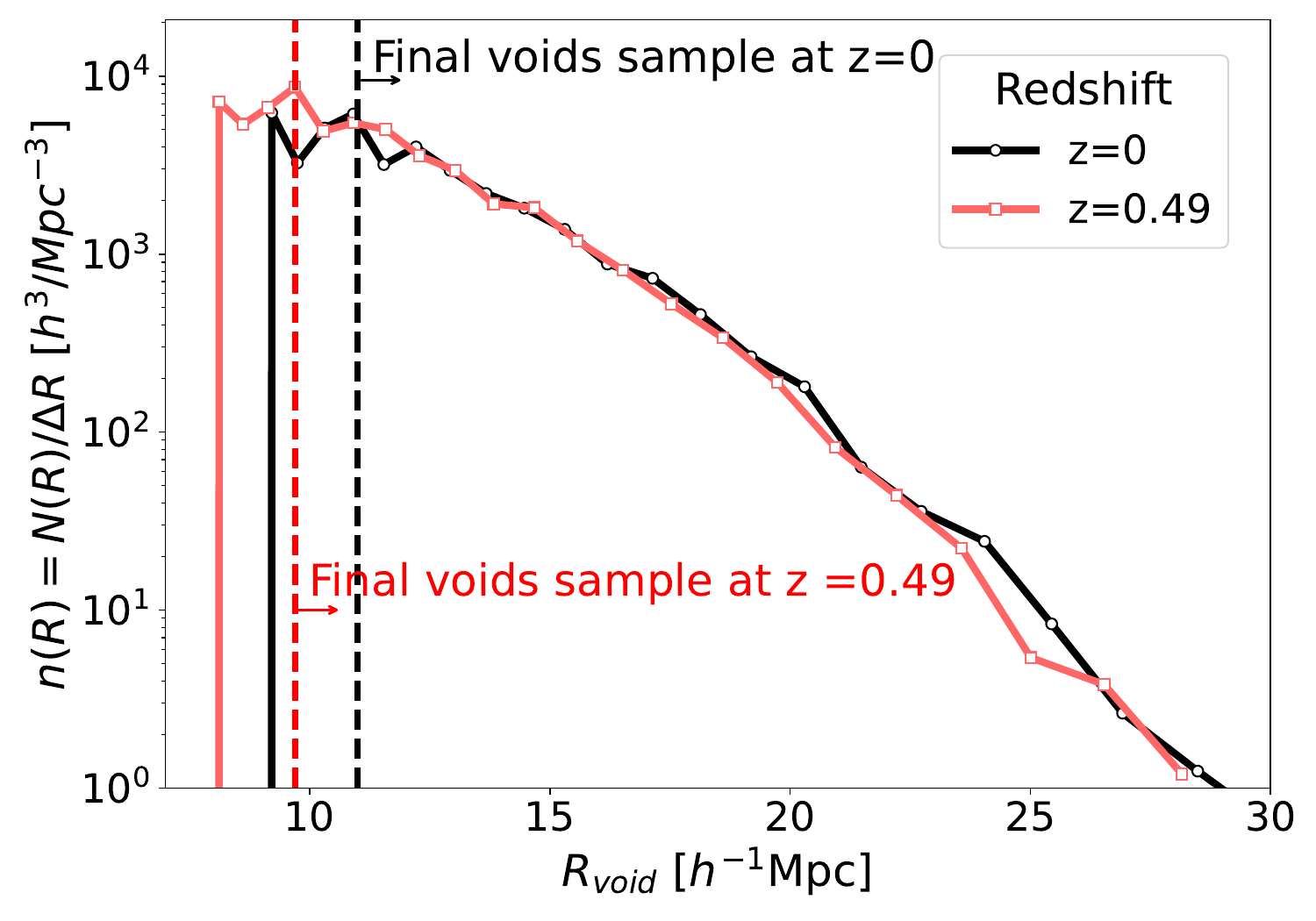}
\end{center}
\caption{\label{fig:rvoid_hist_z0} $z=0$ (black lines) and $z=0.49$ (red lines) void size distribution. 
The dashed vertical lines mark the radius up to which the sample can be considered complete and which 
make up the final sample of objects.}
\end{figure}

We identify spherical voids in each snapshot of the MDPL2-SAG catalog using the public algorithm
\textsc{Sparkling}\footnote{\url{https://gitlab.com/andresruiz/Sparkling}} \citep{ruiz_void_2015,ruiz_into_2019}. 
This is the same algorithm used in \citet{Alfaro2020} and \citet{alfaro_hod_2022}, which allows a direct comparison of the results.
Briefly, the algorithm works as follows:
\begin{itemize}
    \item First, the density field is estimated using a Voronoi tessellation of the galaxy distribution within each snapshot, using galaxies as tracers.
    \item For each Voronoi cell, a density is calculated as the inverse of the cell volume, and a density contrast ($\delta_{\rm cell}$) is defined relative to the mean density of the tracers ($\bar{\rho}$).
    \item All Voronoi cells that satisfy $\delta_{\rm cell} < 0$ are selected as underdense regions and potential void centers.
    \item Around these centers, the spherical regions grow until the integrated density contrast $\Delta(r) = \delta(<r)$ reaches a threshold value $\Delta_{\rm lim}$ on a given scale $r=R_{\rm void}$, defined as the void radius.
    \item The algorithm then performs a sampling procedure to refine the void center location, seeking to maximize the void radius. 
    This involves making random displacements to the center, covering the volume of the corresponding sphere associated to the Voronoi cell, and accepting the new position if the new void radius is larger.
    \item Finally, overlapping void candidates are removed by keeping the largest voids that do not superpose with any other candidate.
\end{itemize}
For each snapshot, we used the brightest galaxies, specifically those with absolute magnitudes brighter than $M_r - 5\log_{10}(h) = -20$, as tracers for the density field.
The voids were defined using a density contrast threshold of $\Delta_{\rm lim} = -0.9$, which means that the identified voids have an integrated density that is, at most, 10\% 
of the mean density of the tracers.
This choice is motivated by the fact that identified voids with this value $\Delta_{\rm lim}$ exhibit the most significant differences in their HODs compared to the general halo population \citep{Alfaro2020}.
By applying this algorithm to each snapshot, we obtain an initial sample of cosmic voids. 
To construct the final void sample, we examined the void radii distribution in each snapshot and identified the minimum radius at which the distribution of the void radii starts to decrease.
This decrease indicates that the void sample is not complete for smaller radii due to resolution effects, something expected taking into account the shot-noise predominance in low density small scales. 
Therefore, for each snapshot, we only considered voids with radii larger than this identified minimum radius.
As an example, Figure \ref{fig:rvoid_hist_z0} shows the void size distributions for snapshots 125 and 107 ($z=0$ and $z=0.49$, respectively), where the vertical dashed lines mark the minimum radius in each case.
%
Taking into account this radius cut in each void sample, we obtain our cosmic void catalogs to study the HOD redshift evolution within these low-density environments. 
The final number of voids identified in each snapshot and the total number of void galaxies are shown in Table \ref{tab:snapshots} .
%
%

\section{Effects of cosmic voids in the HOD across the redshift}
\label{sec:HOD_results}

\subsection{The global effect of the cosmic voids.}
\label{sec:global_effects}

\textcolor{black}{To understand how the global effect of voids on the halo population changes over time, in this section, we analyze void galaxies and halos at different redshifts. It is important to clarify that in this part of the work we are not tracking the same objects across snapshots using merger trees. Instead, we independently identify cosmic voids at each redshift and analyze the properties of galaxies and halos that reside within these voids at each specific epoch.}
We consider snapshots from $z=0$ to $z=0.49$ (snapshots 125 to 107, see Table \ref{tab:snapshots}) of the MDPL2-SAG simulation.
For our samples in each snapshot, we select all halos whose central galaxy is located within the boundaries of a cosmic void.
We then include all the galaxies associated with these halos.
%
%

%
Following the same procedure used in \cite{Alfaro2020}, to determine the HOD within the voids, we perform a stacking analysis of the selected halos in each snapshot.
This allows us to compute the average number of galaxies as a function of halo mass within the void environment, $\langle N|M_{\rm halo} \rangle$.
We then compare this measurement with the overall HOD measured in the entire simulation box. 
This provides us with a clear picture of how the presence of voids affects the halo occupation at different epochs. 
%

\begin{figure*}[h!]
\begin{center}
    \includegraphics[width=\textwidth]{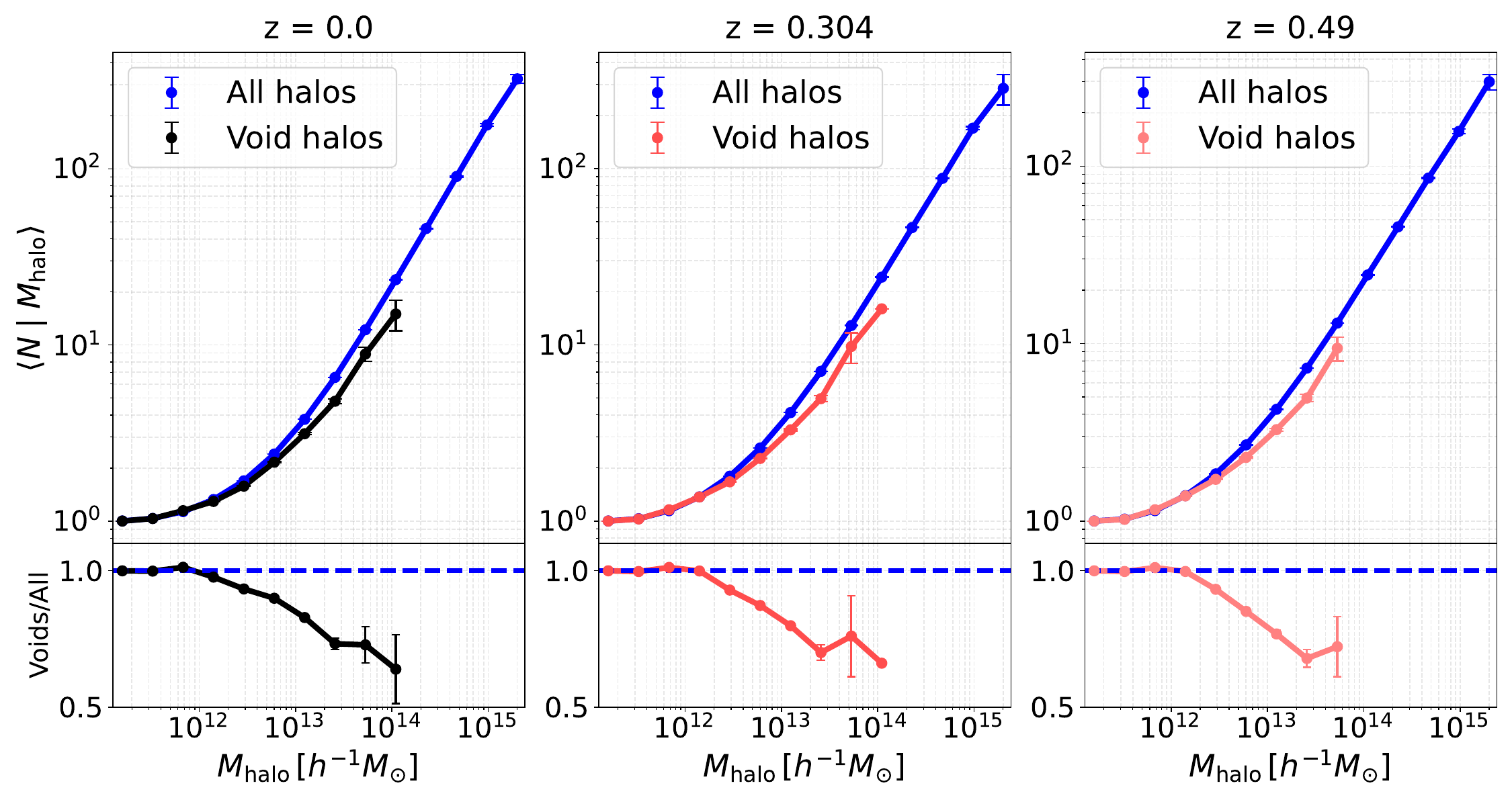}
    \end{center}
    \caption{HOD measured for galaxies with $M_r - 5log_{10}(h) < -17$ at redshifts $z=0$, $z=0.3$, and $z=0.5$ for the MDPL2-SAG simulation. The blue lines represents the HOD for the complete catalogs, while the black and red lines shows the HOD for halos inside cosmic voids identified in each snapshot. The bottom panels shows the ratio of the HOD inside voids to the HOD in the complete catalogs at each redshifts.\textcolor{black}{The error bars was estimated by the Jackknife resampling method using 50 subsamples.}    
    }
    \label{fig:all_vs_voids}
\end{figure*}

%
Figure \ref{fig:all_vs_voids} shows the HOD measurements for galaxies with $M_r - 5\log_{10}(h) < -17$ at redshifts $z=0$, $z=0.304$, and $z=0.49$. 
We note that the calculations were performed for the complete sample of snapshots, but for simplicity, only these three redshifts are displayed.
The same applies to the limiting values of $M_r - 5\log_{10}(h)$, where we explored all the results presented in this work for the magnitudes $M_{r,\rm{lim}} = -16$, $-18$, $-19$, $-20$, and $-21$, consistently finding coherent values.
We conclude that the detected effects impact the overall galaxy population regardless of their magnitude.
Therefore, we choose to present the results only for $M_r - 5\log_{10}(h)< -17$, as this sample contains a large number of well-defined objects that are not near the cutoff values used to define the catalogs.
\textcolor{black}{In this and all subsequent graphs that feature error bars, the bars represent the standard error estimated by the Jackknife resampling method. In each case, we used 50 subsamples, since the jackknife estimates stabilize once at least 30 subsamples are employed.}
In each panel of Figure \ref{fig:all_vs_voids}, we compare the HOD measured for halos inside voids (black line for the $z=0$ snapshot and red lines for higher redshifts, with lighter shades indicating higher redshifts) with the HOD measured for all halos in the corresponding snapshot (blue lines).
The bottom panels show the ratios of the HOD within voids to the overall HOD for the same three redshifts.
It can be seen that the HOD within voids is systematically lower compared to the overall HOD at all redshifts, which agrees with previous results of our previous work showing that the number of galaxies inside voids is lower than in other environments.
In all cases, up to $\sim 10^{12}h^{-1}M_{\odot}$, the halos inside the voids do not show significant differences with the redshift compared to the global result.
However, for halos more massive than $\sim 10^{12}h^{-1}M_{\odot}$, the differences between the HOD in voids and overall increase with the halo mass for all redshifts, which are larger for larger masses.
It is also worth noting that the number of halos comprising the two last mass bin samples decreases when we consider higher redshifts, to the point where the sample corresponding to $z=0.49$ does not reach the last mass bin.
These results show that the effect of the void environment on the HOD is consistent across all analyzed redshifts, suggesting that the mechanisms that cause the HOD suppression within voids are similar through cosmic time.
%

\subsection{The effect of the cosmic voids on $z=0$ void galaxies.}
\label{sec:galaxies_effects}

\begin{figure}[h!]
\begin{center}
\includegraphics[width=\columnwidth]{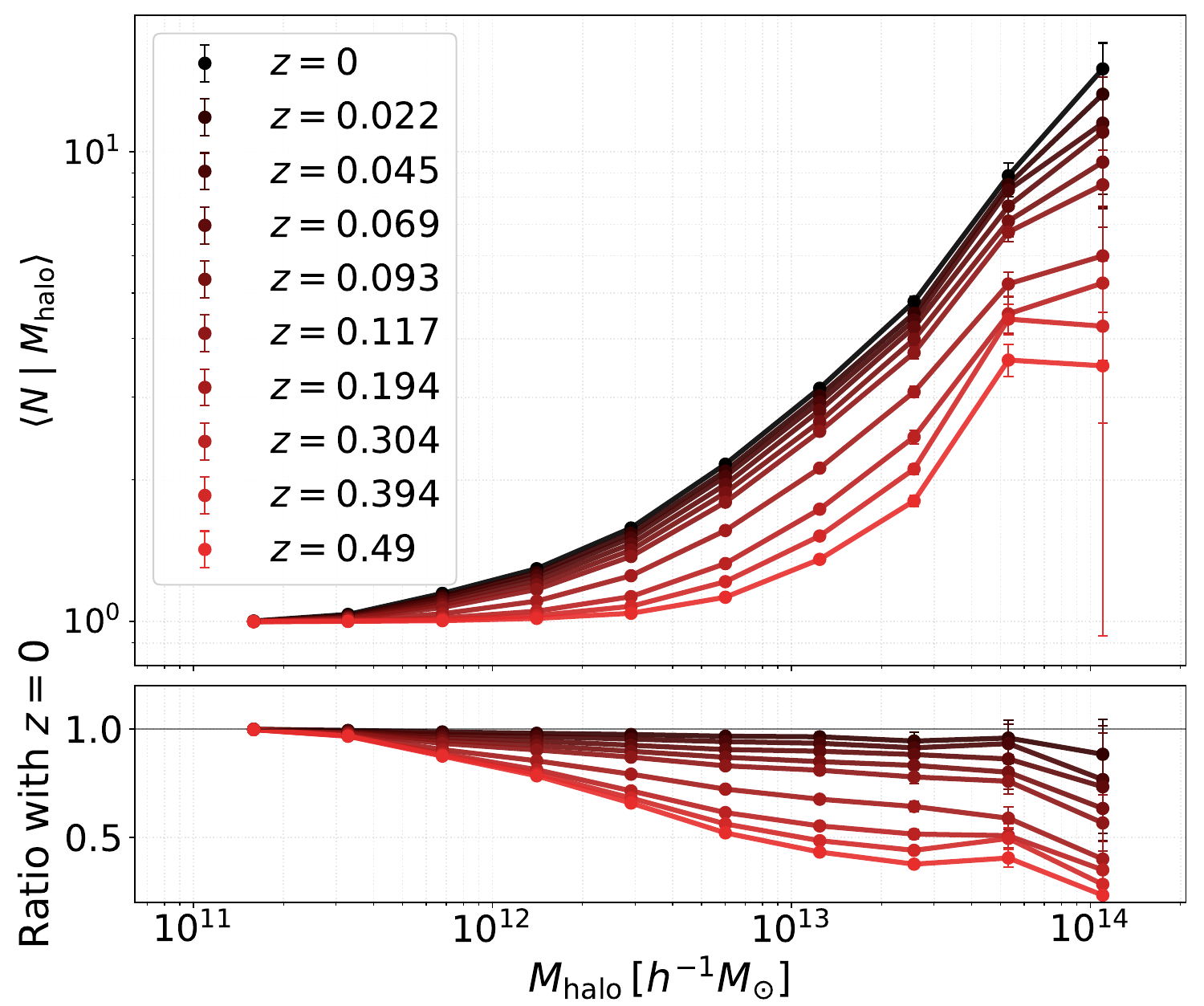}
\end{center}
\caption{\label{fig:hod_central_z0} Evolution of the HOD of central galaxies in voids at $z=0$. Top panel: HOD for central galaxies in voids at $z=0$ across different redshifts. Bottom panel: ratio between the HOD at each redshift and the HOD at $z=0$. \textcolor{black}{The error bars was estimated by the Jackknife resampling method using 50 subsamples.}}
\end{figure}

%
In the previous section, we analyzed the global effect of voids on the HOD, finding that this global effect persists with redshift. 
Now, we will focus on how the voids influence the galaxy population residing within them at $z=0$, and trace the evolution of this particular population over time.
To examine the impact of voids on the HOD of their member galaxies, we track the evolution of central galaxies identified within voids at $z=0$. 
This is achieved using the \texttt{GalaxyStaticID} parameter from the MDPL2-SAG catalog, which remains static across all simulation snapshots and allows for consistent galaxy tracking throughout cosmic time. 
Unlike the previous section, where the HOD of halos was analyzed independently at each redshift, this approach focuses on a fixed population of galaxies traced across snapshots.
For each snapshot, we compute the HOD of halos hosting central galaxies that were identified within voids at $z=0$ and remain central in subsequent snapshots. 
Fixing $M_{\rm halo}$ and $M_r - 5\log_{10}(h)$ to their $z=0$ values, we isolate the HOD evolution from effects such as halo mass accretion and stellar evolution. 
This methodology provides a consistent view of how the HOD evolves over time for a single, defined galaxy population.
%

\begin{figure}[h!]
\begin{center}
\includegraphics[width=\columnwidth]{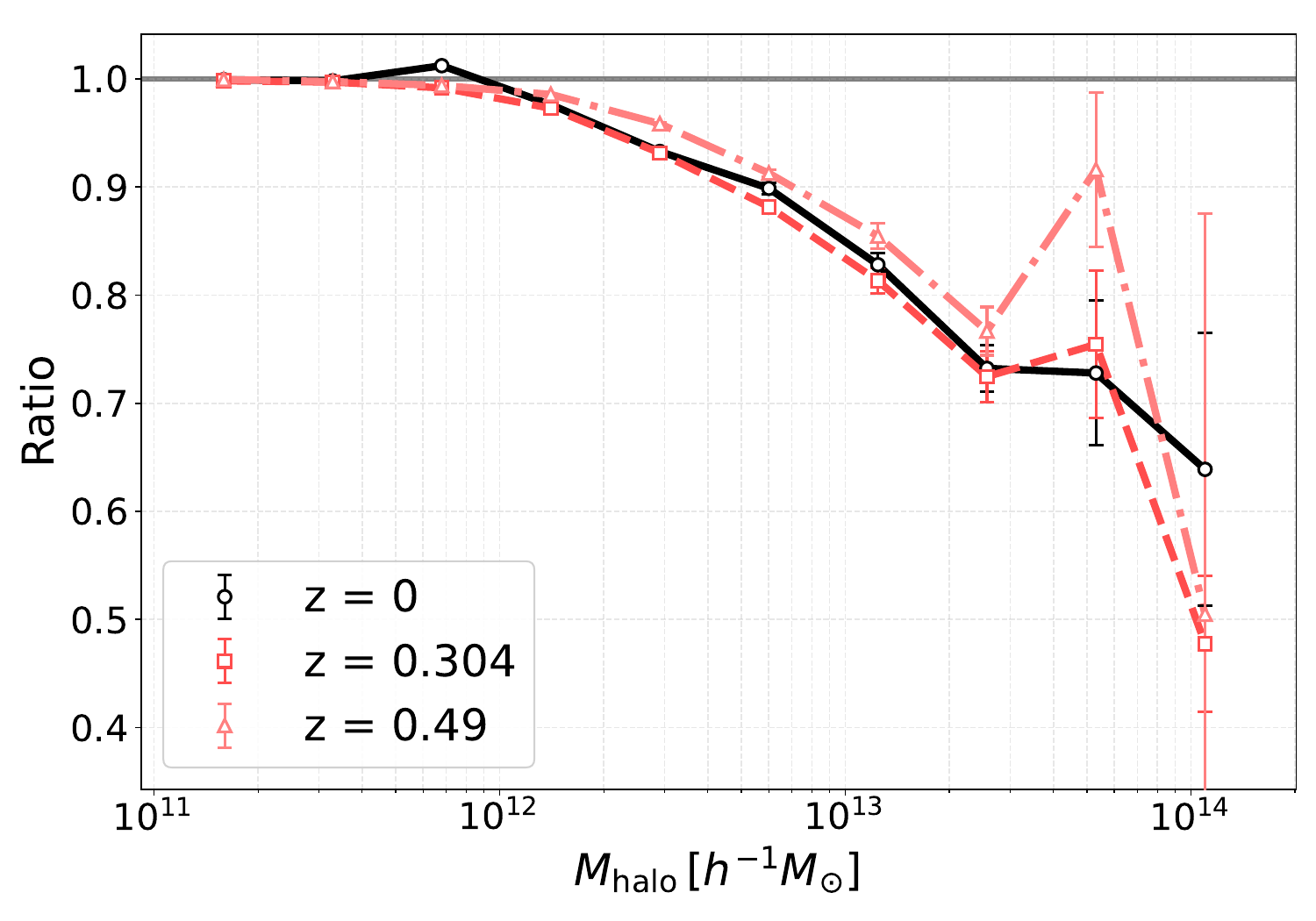}
\end{center}
\caption{\label{fig:ratio_z0} \textcolor{black}{Ratio between the HOD of central galaxies located in voids at z = 0 and the general HOD of the universe at three different redshifts. The black solid line with diamonds corresponds to $z = 0$, the red dashed line with squares to $z = 0.304$, and the pink dashed line with triangles to $z = 0.49$. The HOD for all cases is calculated for galaxies with $M_r - 5log_{10}(h) < -17$ at $z=0$, while fixing halo masses to their $z = 0$ values. The error bars was estimated by the Jackknife resampling method using 50 subsamples.}
}
\end{figure}

%
Figure \ref{fig:hod_central_z0} shows the HOD of central galaxies in voids at $z=0$ across different redshifts for galaxies with $M_r - 5\log_{10}(h) < -17$. 
The top panel presents the HOD curves in the ten selected snapshots between $z=0$ and $z=0.49$, using a black line for the sample at $z=0$ and different shades of red that become lighter at higher redshifts.
%
%
The bottom panel shows the relationship between the HOD at each redshift and the HOD at $z=0$, highlighting the relative changes over time.
%
\begin{figure*}[h!]
 \begin{center}
 \includegraphics[width=\textwidth]{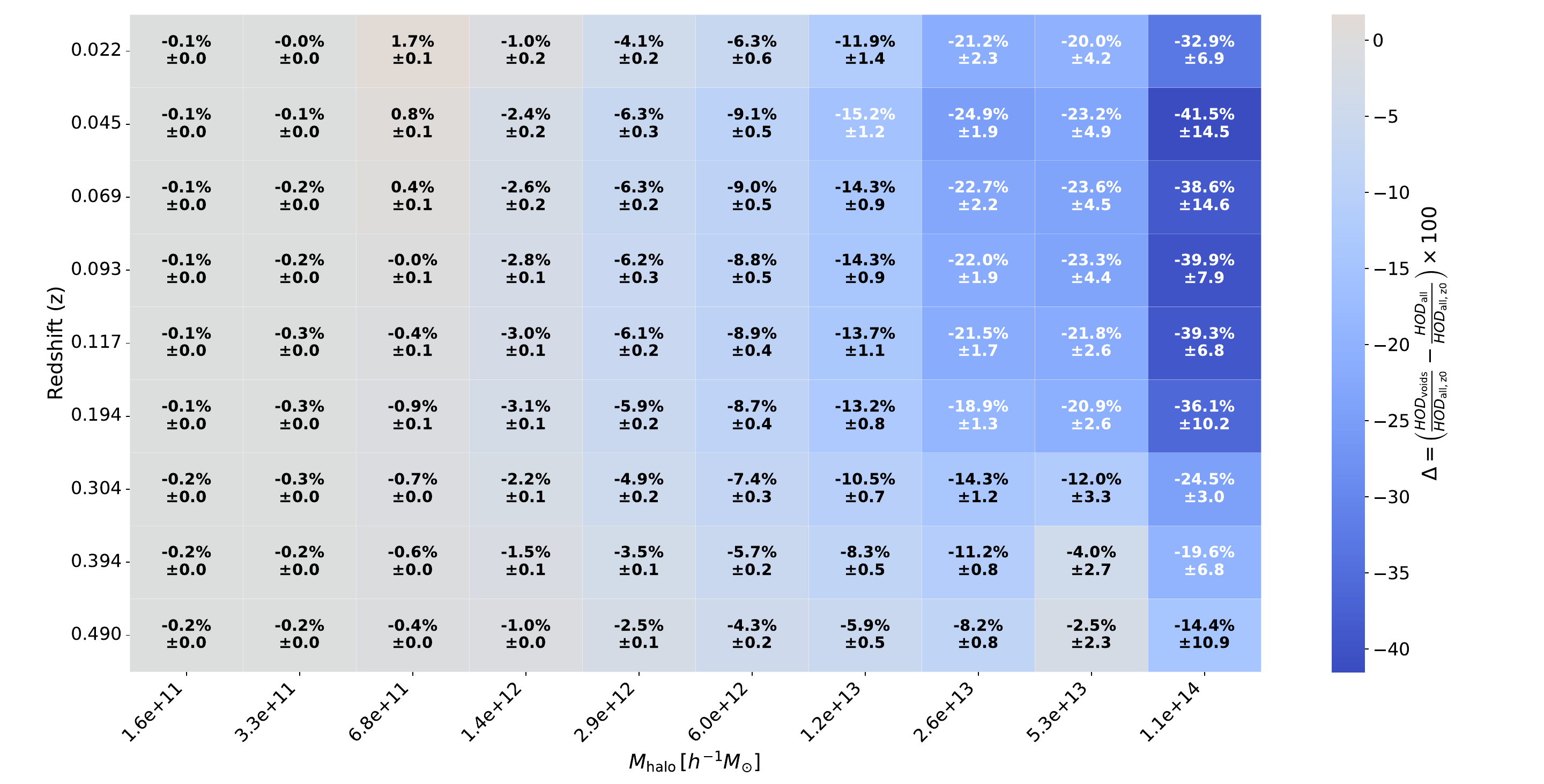}
 \end{center}
 \caption{Variation of the HOD within voids and in the complete snapshot between each redshift and the HOD at $z=0$. \textcolor{black}{The errors was estimated by the Jackknife resampling method using 50 subsamples.}}
 \label{fig:Delta}
\end{figure*}
%
The figure reveals distinct trends depending on the halo mass. 
For halos with $M_{halo} \lesssim 2 \times 10^{11} h^{-1} M_\odot$, the HOD remains nearly constant across all redshifts, indicating that these low-mass halos do not significantly accrete satellite galaxies over time.
However, beyond this mass threshold, the HOD curves start to grow as the redshift decreases, reflecting the progressive accretion of satellite galaxies. 
This growth becomes more pronounced for halos with larger masses, highlighting the increasing efficiency of massive halos in the capture of satellites over cosmic time.
\textcolor{black}{This is a consequence of the hierarchical structure formation scenario}, where massive halos grow predominantly through mergers and accretion at later times \citep{Cooray2002}.
To compare the HOD of central galaxies in voids with the general behavior of halos, we compute the overall HOD by considering the complete box of each snapshot.
We selected all central galaxies at $z=0$ that remain as central in each snapshot, fixed their $M_{\rm halo}$ and $M_r - 5\log_{10}(h)$ values to their $z=0$ properties, and calculated the HOD for galaxies with $M_r - 5\log_{10}(h) < -17$.
Figure \ref{fig:ratio_z0} shows the relationship between the HODs in the previous figure and the general HOD at each redshift.
To avoid line overlap and highlight the main trends, we present only the results for three snapshots: the black solid line with circles corresponds to $z=0$, the red dashed line with squares to $z=0.304$, and the pink dashed line with triangles to $z=0.49$.
Figure \ref{fig:ratio_z0} shows that, in general, galaxies residing in voids at $z=0$ maintain a lower HOD than the average across the entire redshift range studied.
No systematic evolution is observed in the relationship between the HOD of this particular sample of central galaxies and the general trend.
Finally, Figure \ref{fig:Delta} illustrates the relative evolution of the HOD in cosmic voids compared to the general universe in different redshift and halo mass ranges.  
We quantify this evolution through a differential metric ($\Delta$) that compares the normalized HOD in the voids against the cosmic mean, using $z=0$ as the reference epoch.  
The differential metric is computed as:
\begin{multline}
\Delta = \left[ \left(1 - \frac{HOD_{\mathrm{all}}}{HOD_{\mathrm{all, z0}}}\right) - \left(1 - \frac{HOD_{\mathrm{voids}}}{HOD_{\mathrm{all, z0}}}\right)\right] \times 100 \\
= \left(\frac{HOD_{\mathrm{voids}}}{HOD_{\mathrm{all, z0}}} - \frac{HOD_{\mathrm{all}}}{HOD_{\mathrm{all, z0}}}\right) \times 100
\end{multline}
where each term in parentheses on the first line of the equation represents the corresponding percentage with respect to the HOD value at $z=0$ for each mass bin.  
This normalization allows us to track relative changes in the galaxy occupation of halos while accounting for the natural evolution of the overall HOD.  
The uncertainties were derived through jackknife resampling and proper error propagation.  
The heatmap reveals several notable trends. First, a clear mass-dependent evolution emerges, with stronger differential effects at higher halo masses across all redshifts.  
This finding further supports the idea that the void environment exerts a more significant influence on the galaxy occupation of massive halos.  
Secondly, a redshift evolution is evident, where the magnitude of $\Delta$ generally increases with redshift for halos with $M_{\mathrm{halo}} \gtrsim 10^{12} h^{-1} M_\odot$, but remains practically unchanged for less massive halos. 
This indicates a growing disparity in galaxy occupation patterns between low- and high-mass halos, driven by the void environment as we look back in time.  
%
%

\section{Void galaxies at $z=0$: formation and evolution of the local environment}
\label{sec:mass_analysis}

Building on the previous section's investigation of the impact of voids on the HOD and the properties of galaxy populations whose central galaxy resides within a void at $z=0$, this section offers a more detailed examination of their characteristics.
Our goal is to understand how these galaxies have been shaped by the void environment and to explore the underlying reasons for their significantly lower HODs compared to the broader galaxy population.
%

\begin{figure}[h!]
\begin{center}
\includegraphics[width=\columnwidth]{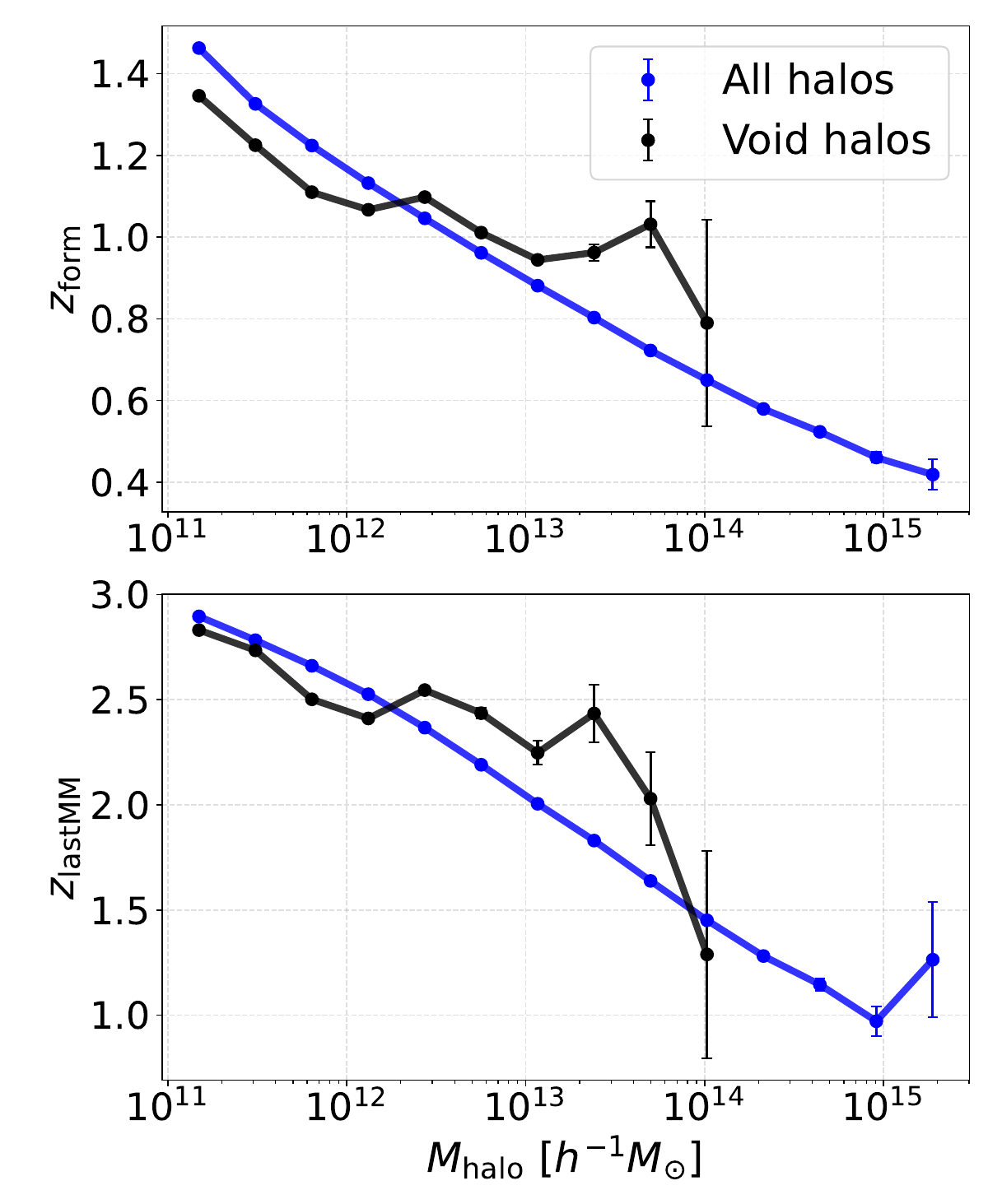}
\end{center}
\caption{\label{fig:zform} Halos formation redshifts in function of its mass.The top panel shows the results for $z_{\text{form}}$, defined as the redshift when half of the maximum halo mass was accreted. The bottom panel correspond to $z_{\text{lastmm}}$, the redshift at which the last major merger occurred. The blue lines represent the complete $z=0$ snapshot and the black line the $z=0$ void halos samples. \textcolor{black}{The error bars was estimated by the Jackknife resampling method using 50 subsamples.}}    
\end{figure}

%
We begin our analysis by investigating the formation times of halos hosting central galaxies within voids at $z=0$.
This step is essential for understand the evolutionary history of these systems and how their development might be influenced by the void environment.
For halos within voids, we compute two key formation metrics: formation redshift ($z_{\text{form}}$), defined as the redshift when half of the maximum halo mass was accreted, and the redshift of the last major merger ($z_{\text{lastMM}}$), defined as the redshift at which the last major merger occurred (mass ratio > 0.3). 
Both parameters are readily inferred from the \texttt{halfmass\_scale} and \texttt{lastmm\_scale} outputs provided by the \textsc{Rockstar} algorithm.
Figure \ref{fig:zform} presents these results, showing the formation times as a function of the halo mass. 
The top panel displays the \textcolor{black}{mean values} of $z_{\text{form}}$, while the bot panel illustrates the distribution of $z_{\text{lastmm}}$.
A clear trend emerges from these results, the less massive halos within voids are, on average, younger than their counterparts in the general population, whereas the most massive halos in voids are older than the average. 
This finding, when correlated with the HOD results from the previous section, suggests that the void environment exerts distinct influences on halos of different masses. Additionally, the formation history of halos appears to play a crucial role in shaping the galaxy populations they host.
\textcolor{black}{
These observed trends in halo formation times and merger histories align with recent studies of galaxies in underdense environments. For example, \cite{Rosas_Guevara_2022} identified analogous mass‐dependent patterns in the timing of merger events across different cosmic regions using hydrodynamical simulations, noting that major mergers in massive-void galaxies tend to occur more recently toward the inner void than in its outskirts. Similarly, \citep{Rodriguez-Medrano2023} found that in the last 2 Gyr, the fraction of accreted mass from high‐mass halos in voids has declined relative to low‐mass halos, suggesting that more massive systems in these regions may have reached a dynamically quieter state. Our results corroborate these findings and extend them by elucidating how such formation histories specifically shape the halo occupation distribution within void regions. Building on this framework, we proceed to investigate the spatial arrangement and evolutionary pathways of halo densities in void environments.
}
%

\begin{table}
\centering
\begin{tabular}{ccc}
\hline\hline
Snapshot & Redshift & $D_5 = 1$ scale \\ 
& & ($\mathrm{Mpc}\,h^{-1}$) \\
\hline
125 & 0.000 & 5.17 \\
124 & 0.022 & 5.13 \\
123 & 0.045 & 5.09 \\
122 & 0.069 & 5.05 \\
121 & 0.093 & 5.01 \\
120 & 0.117 & 4.98 \\
117 & 0.194 & 4.89 \\
113 & 0.304 & 4.81 \\
110 & 0.394 & 4.78 \\
107 & 0.490 & 4.80 \\
\hline
\end{tabular}
\caption{
\textcolor{black}{
Mean local density scale of the selected snapshots used in this work. Column 1: snapshot number from the MDPL2-SAG simulation. Column 2: corresponding redshift. Column 3: Average distance from central galaxies to the fifth nearest bright neighbor galaxy with $M_r - 5\log_{10}(h) < -20$, this scale corresponds to $D_5 = 1$ value.}
}
\label{tab:snapshots_d5}
\end{table}

\begin{figure*}[h]
\begin{center}
\includegraphics[width=\textwidth]{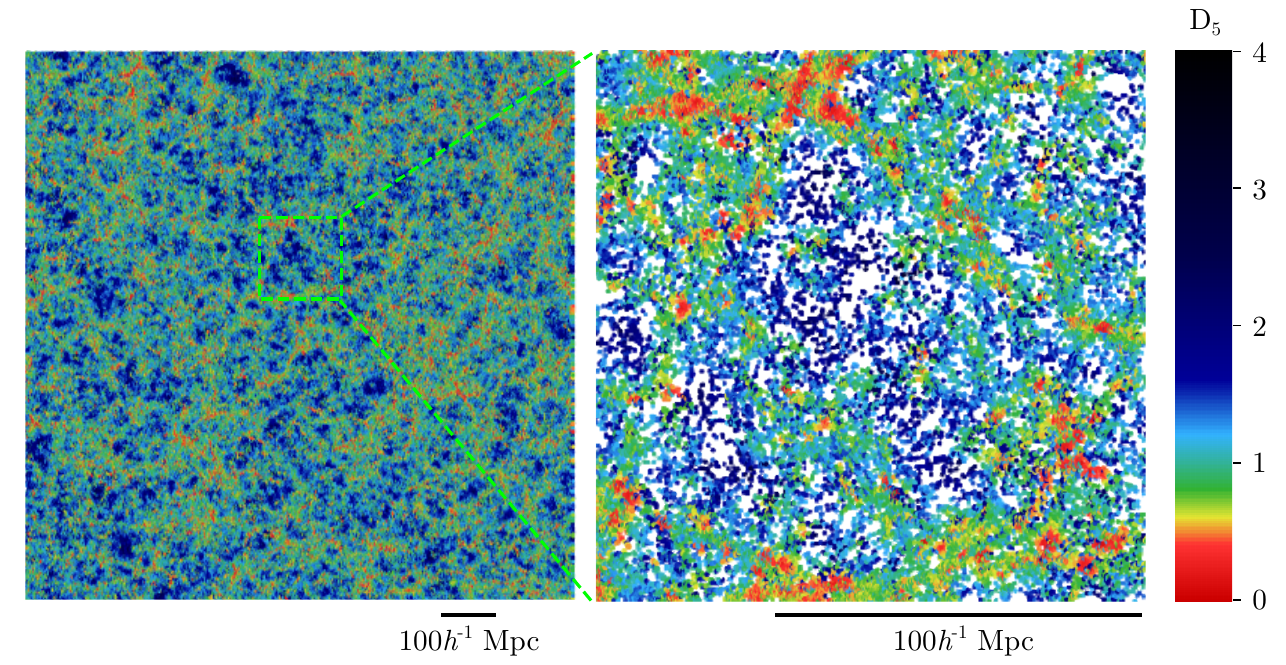}
\end{center}
\caption{\label{fig:Scatter} Spatial distribution of central galaxies at $z=0$ within a simulation box slice spanning $100\,\mathrm{Mpc}\,h^{-1} < z < 150\,\mathrm{Mpc}\,h^{-1}$. The color of each point represents $D_5$, the distance to the fifth nearest bright neighbor, normalized by the mean value of the complete snapshot, serving as a proxy for the local galaxy density. The \textcolor{black}{left} panel displays an area of $1000\,\mathrm{Mpc}\,h^{-1}$ per side, while the \textcolor{black}{right} panel presents a zoom-in of the highlighted region.}
\end{figure*}

%
To quantify the local density, we measure the distance from each central galaxy to its fifth nearest neighboring galaxy brighter than $M_r - 5\log_{10}(h) = -20$, regardless of whether the neighbor is a central or satellite galaxy, and normalized this quantity by the average fifth nearest bright neighboring of the complete snapshot.
This parameter, referred as $D_5$, provides a reliable measure of the local galaxy density surrounding each halo.
A value of $D_5$ $< 1$ indicates a high local density, while the values $> 1$ suggest a more isolated environment. 
\textcolor{black}{The choice of $M_r - 5\log_{10}(h) = -20$ as a brightness threshold in each snapshot is consistent with our void identification methodology, ensuring the use of the same tracers of the distribution of the underlying field matter for environmental characterization. Table \ref{tab:snapshots_d5} shows the physical scale corresponding to $D_5=1$ in each snapshot, which typically represents environments with densities comparable to the mean cosmic density. As expected, the scales of $D_5=1$ tend to increase as we approach $z=0$. }
Figure \ref{fig:Scatter} illustrates the spatial distribution of central galaxies at $z=0$, showing a slice of the simulation box with a thickness of $50 \hmpc$. 
Each central galaxy is represented by a point, colored according to its $D_5$, providing an intuitive visualization of density variations throughout the large-scale structure.
%

\begin{figure*}[h]
\begin{center}
\includegraphics[width=0.95\textwidth]{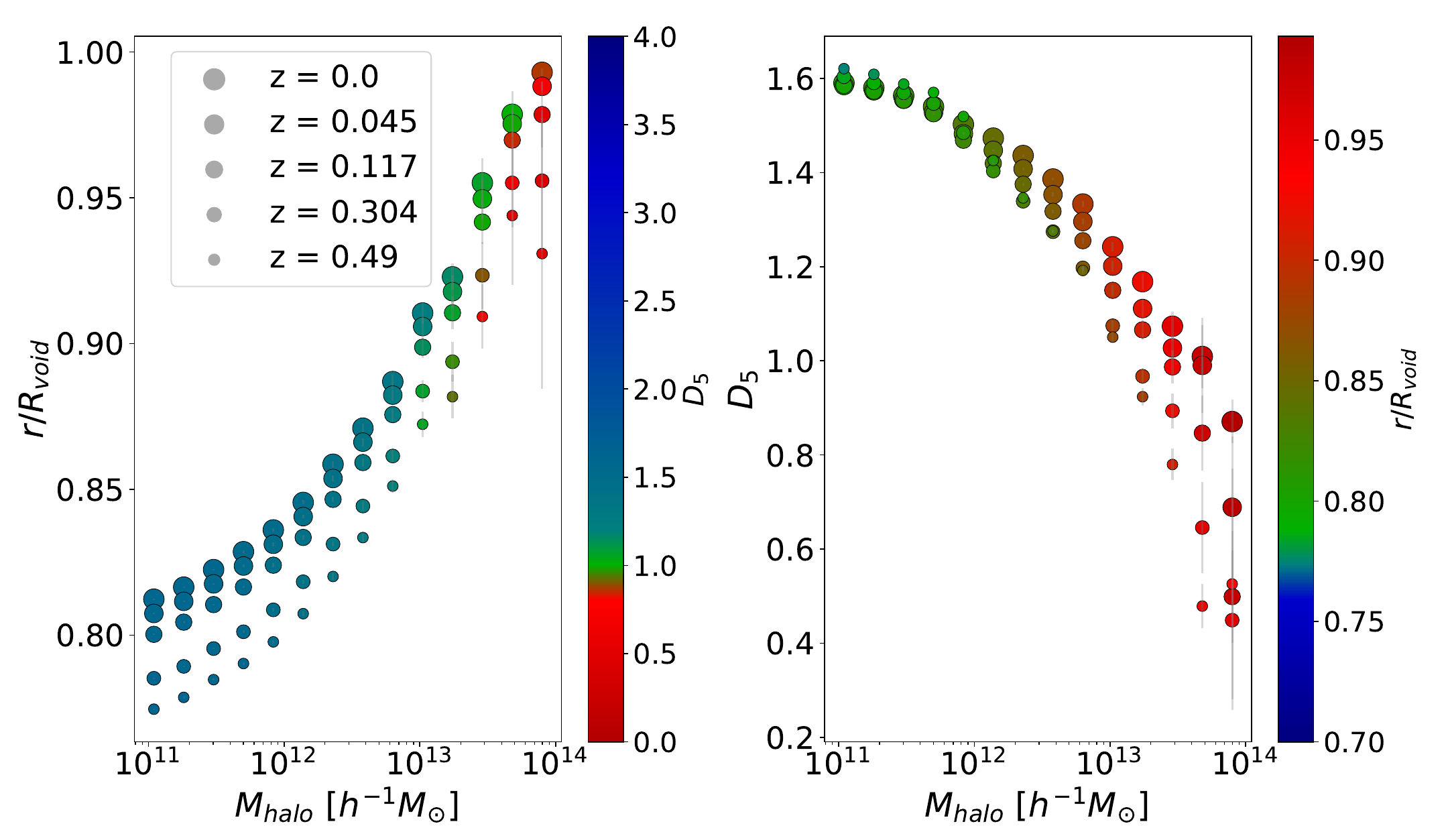}
\end{center}
\caption{\label{fig:rvoid_distance} Left panel: \textcolor{black}{Normalized distance from the void center of central galaxies in voids at $z=0$ as a function of their halo mass for five redshifts samples. Circle sizes decrease with increasing redshift, and the color scale represents the mean $D_5$ for each mass bin}. Right panel: Variation of $D_5$ with halo mass for the same samples, where the color scale indicates the mean $r/R_{\rm{void}}$ for each bin. \textcolor{black}{The error lines was estimated by the Jackknife resampling method using 50 subsamples.}} 
\end{figure*}

%
In addition to these parameters, we analyze the spatial distribution of these central galaxies at each redshift.  
For each galaxy, we compute the distance $r$ from the center of the void it will inhabit at $z=0$.  
In this way, $r/R_{\rm{void}}$ serves as an indicator of the galaxy's location within the region that will eventually evolve into a void at $z=0$.
Thus, the combined analysis of $r/R_{\rm{void}}$  and $D_5$ allows us to determine whether these galaxies have always resided in the inner regions of the area that will evolve into a void at $z=0$.  
In addition, it enables us to quantify how much the local density surrounding them has changed as this region evolved.
Figure \ref{fig:rvoid_distance} visualizes these results. 
The left panel shows the relationship between the halo mass of central galaxies in voids at $z=0$ and their normalized distance from the center of the void for five of the redshifts studied.
These are represented by circles of different sizes, which decrease as the redshift increases. 
The color scale associated with each point corresponds to the mean $D_5$ for each mass bin.
The right panel, on the other hand, illustrates how $D_5$ varies with the mass of these same halos.
In this case, the color scale represents the mean value of $r/R_{\mathrm{void}}$ for each sample.
The combined analysis of both panels reveals how the halos in each mass bin evolve within the region that will form the void at $z=0$.
We observe a clear trend indicating a correlation between halo mass and their location within the void.
Lower-mass halos tend to populate the innermost regions, while more massive halos are found closer to the void walls.
\textcolor{black}{This mass segregation within voids is consistent with previous findings by \cite{Gottlober_2003,Ricciardelli_2014,Ricciardelli_2017}} for galaxies.
Additionally, as expected, the local density of halos increases as we move toward the outer regions of the void, leading to a direct correlation between the halo mass and its local density.
%

%
Examining the evolution of these results with redshift, we find that, in general, galaxies that populate a void at $z=0$ have always resided within the volume that defines these regions. 
The overall trend shows that these galaxies migrate from more central areas toward the outer parts of the void.
\textcolor{black}{
This behavior aligns with the well-established dynamics of cosmic voids. Previous studies have demonstrated that voids behave similarly to underdense regions in an expanding universe, where the inner shells expand faster than the outer ones, creating a distinct velocity field characterized by coherent outflows \citep{Dubinski1993,padilla_void_2005,ceccarelli_voids_2006}. This differential expansion causes galaxies within voids to gradually migrate toward the boundaries over cosmic time, consistent with our findings.
}
At the same time, we see that the evolution of local density differs depending on the halo mass range, following a pattern similar to that observed in the HOD.
Halos with $M_{\mathrm{halo}} \lesssim 10^{12} h^{-1} M_\odot$ show no significant variations in $D_5$.
However, for halos with $M_{\mathrm{halo}} \gtrsim 10^{12} h^{-1} M_\odot$, it becomes evident that the local density decreases as the redshift approaches zero. 
This effect is increasingly pronounced for more massive halos: halos with $M_{\mathrm{halo}} \gtrsim 10^{13} h^{-1} M_\odot$ transition from residing in locally overdense regions \textcolor{black}{($D_5<1$)} to being surrounded by locally underdense environments \textcolor{black}{($D_5>1$)}.
In particular, the local density of halos in the highest mass bin remains above 1 even at $z=0$, reinforcing the idea that these halos lie near the void boundaries, at the interface with higher-density structures such as sheets or filaments.
These findings are consistent with a scenario in which the voids are populated by halos with distinct formation histories driven by their halo masses, as we describe in Figure \ref{fig:zform}.
Low-mass halos are younger than average and predominantly occupy the inner regions of voids.
Most of these halos have inhabited void interiors since high redshifts, evolving in sub-dense regions.
In contrast, more massive void halos tend to be located on void walls and are older than average.
Most of these halos transition from regions of high local density to sub-dense and average environments, potentially impacting their development.
These differences in void halo populations may explain the variations detected in the HOD within these regions.
%

\section{Summary and conclusions}
\label{sec:conclusions}

In this paper, we have investigated the evolution of the halo occupation distribution (HOD) within cosmic voids from redshift $z=0.5$ to the present day, utilizing the MDPL2-SAG semi-analytic galaxy catalog and the \textsc{Sparkling} void finder algorithm.
Our study aimed to understand how the underdense void environment influences the HOD over cosmic time and to explore the underlying mechanisms responsible for the distinctive galaxy populations observed in these regions.
Our analysis of the HOD across different redshifts revealed several key findings:
\begin{itemize}
    \item {\bf Persistent suppression of the HOD in voids:} We confirm that the HOD within cosmic voids is consistently lower than the global HOD across the redshift range studied ($z=0$ to $z=0.49$). This suppression is evident for halos with masses greater than $\sim 10^{12}h^{-1}M_{\odot}$. This indicates that the void environment consistently hinders galaxy occupation in halos, especially in more massive systems, throughout cosmic history.

    \item {\bf Stable relative HOD for $z=0$ void galaxies:} When tracing the HOD evolution of galaxies that reside in voids at $z=0$ (and were central galaxies over time), we found that their HOD remains consistently lower than the global HOD at all redshifts.  Importantly, the ratio of their HOD to the global HOD does not show significant redshift evolution. This implies that the factors that determined the lower HOD of these specific void galaxies were already in place at higher redshifts and maintained a consistent relative effect over time.

    \item {\bf Mass-dependent environment in voids:}  
    The spatial distribution of void halos at $z=0$ exhibits a clear mass dependence. Lower-mass halos tend to reside in the inner regions of voids, whereas more massive halos are preferentially found near the void boundaries, \textcolor{black}{in agreement with previous studies \citep{Gottlober_2003, Ricciardelli_2014,Ricciardelli_2017}}. Moreover, the local density around massive void halos decreases significantly from high redshifts to $z=0$, indicating a transition from initially denser environments to underdense void regions. In contrast, lower-mass halos consistently residing in underdense environments since early times.

    \item {\bf Mass-dependent halo formation:}
    The particular environmental conditions within voids create a relatively quiescent setting for halos, where mergers and interactions with other structures are less frequent compared to higher-density regions. Low-mass halos, which predominantly inhabit locally underdense regions, may have experienced inhibited matter accretion, delaying their assembly compared to halos of the same mass in an average-density environment. Conversely, massive halos have transitioned from initially high-density regions to medium- or low-density void environments, which may have facilitated their earlier assembly compared to similar halos in other cosmic environments.    
    
\end{itemize}

These results suggest a scenario where the void environment exerts a complex, mass-dependent influence on galaxy formation and halo occupation.
The persistently lower HOD in voids across redshift points to fundamental environmental mechanisms at play that suppress galaxy formation within these underdense regions. 
The distinct formation histories and environmental evolution of low- and high-mass halos in voids likely contribute to the observed mass-dependent HOD trends. 
{The mass scale of $\sim 10^{12}h^{-1}M_{\odot}$ emerges as a natural threshold for the global void effects on the HOD, the specific suppression of the HOD for $z=0$ void galaxies, the formation time of void halos, and the local density evolution of $z=0$ void galaxies.
Below this mass, the void environment appears to have minimal impact on all these effects, suggesting that galaxy formation in these lower-mass halos is governed by processes largely independent of the void environment itself.
Above this mass, the void environment becomes a significant factor, suppressing galaxy occupation, particularly in more massive halos.
In the latter, their location on void walls and their transition from denser to underdense environments might influence their accretion history and satellite galaxy capture, leading to observed HOD suppression, particularly at later times.
The distinct evolutionary paths of low- and high-mass halos within voids, in terms of formation time and environmental history, likely underpin these mass-dependent HOD variations.
Future work in hydrodynamical simulations is crucial to disentangle the specific physical mechanisms driving the mass-dependent impact of voids on the HOD, including gas accretion, feedback, and mergers, while also exploring broader environmental factors like tidal forces and merger rates, and incorporating analysis of galaxy properties and their evolution across cosmic time to fully understand the interplay of environment, halo formation history, and galaxy occupation in these underdense regions.
The results in this work present a challenge in contrast to observations.
Therefore, it would be interesting to develop methodologies that allow us to estimate the HOD precisely at different redshifts using spectroscopic surveys or with photometric redshifts that will be available in the upcoming years.
%

\begin{acknowledgements}
The authors thank Dra. Yamila Yaryura for her help with the MDPL2-SAG data. 
The \textsc{CosmoSim} database used in this paper is a service of the Leibniz Institute for Astrophysics Potsdam (AIP).
The \textsc{MultiDark} database was developed in cooperation with the Spanish MultiDark Consolider Project CSD2009-00064.
The authors gratefully acknowledge the Gauss Centre for Supercomputing e.V. (\url{www.gauss-centre.eu}) and the Partnership
for Advanced Supercomputing in Europe (PRACE, \url{www.prace-ri.eu}) for funding the MultiDark simulation project by providing
computing time on the GCS Supercomputer SuperMUC at the Leibniz Supercomputing Centre (LRZ, \url{www.lrz.de}).
Figures were developed using \textsc{Matplotlib} \citep{Hunter2007} and some of them were post-processed with \textsc{Inkscape} (\url{https://inkscape.org})

\end{acknowledgements}

\bibliographystyle{aa}
\bibliography{references}

\begin{thebibliography}{48}
\expandafter\ifx\csname natexlab\endcsname\relax\def\natexlab#1{#1}\fi

\bibitem[{Alam {et~al.}(2015)Alam, Albareti, Prieto, Anders, Anderson, Anderton, Andrews, Armengaud, Aubourg, Bailey, {et~al.}}]{Alam2015}
Alam, S., Albareti, F.~D., Prieto, C.~A., {et~al.} 2015, The Astrophysical Journal Supplement Series, 219, 12

\bibitem[{Alfaro {et~al.}(2020)Alfaro, Rodriguez, Ruiz, \& Lambas}]{Alfaro2020}
Alfaro, I.~G., Rodriguez, F., Ruiz, A.~N., \& Lambas, D.~G. 2020, \aap, 638, A60

\bibitem[{{Alfaro} {et~al.}(2022){Alfaro}, {Rodriguez}, {Ruiz}, {Luparello}, \& {Lambas}}]{alfaro_hod_2022}
{Alfaro}, I.~G., {Rodriguez}, F., {Ruiz}, A.~N., {Luparello}, H.~E., \& {Lambas}, D.~G. 2022, \aap, 665, A44

\bibitem[{{Alfaro} {et~al.}(2021){Alfaro}, {Ruiz}, {Luparello}, {Rodriguez}, \& {Garcia Lambas}}]{Alfaro2021}
{Alfaro}, I.~G., {Ruiz}, A.~N., {Luparello}, H.~E., {Rodriguez}, F., \& {Garcia Lambas}, D. 2021, \aap, 654, A62

\bibitem[{Artale {et~al.}(2018)Artale, Zehavi, Contreras, \& Norberg}]{Artale2018}
Artale, M.~C., Zehavi, I., Contreras, S., \& Norberg, P. 2018, \mnras, 480, 3978

\bibitem[{{Behroozi} {et~al.}(2013{\natexlab{a}}){Behroozi}, {Wechsler}, \& {Wu}}]{behroozi_rockstar_2013}
{Behroozi}, P.~S., {Wechsler}, R.~H., \& {Wu}, H.-Y. 2013{\natexlab{a}}, \apj, 762, 109

\bibitem[{{Behroozi} {et~al.}(2013{\natexlab{b}}){Behroozi}, {Wechsler}, {Wu}, {Busha}, {Klypin}, \& {Primack}}]{behroozi_trees_2013}
{Behroozi}, P.~S., {Wechsler}, R.~H., {Wu}, H.-Y., {et~al.} 2013{\natexlab{b}}, \apj, 763, 18

\bibitem[{{Berlind} \& {Weinberg}(2002)}]{Berlind2002}
{Berlind}, A.~A. \& {Weinberg}, D.~H. 2002, \apj, 575, 587

\bibitem[{Berlind {et~al.}(2003)Berlind, Weinberg, Benson, Baugh, Cole, Dav{\'e}, Frenk, Jenkins, Katz, \& Lacey}]{Berlind2003}
Berlind, A.~A., Weinberg, D.~H., Benson, A.~J., {et~al.} 2003, \apj, 593, 1

\bibitem[{{Ceccarelli} {et~al.}(2008){Ceccarelli}, {Padilla}, \& {Lambas}}]{Ceccarelli2008}
{Ceccarelli}, L., {Padilla}, N., \& {Lambas}, D.~G. 2008, \mnras, 390, L9

\bibitem[{{Ceccarelli} {et~al.}(2006){Ceccarelli}, {Padilla}, {Valotto}, \& {Lambas}}]{ceccarelli_voids_2006}
{Ceccarelli}, L., {Padilla}, N.~D., {Valotto}, C., \& {Lambas}, D.~G. 2006, \mnras, 373, 1440

\bibitem[{{Cooray} \& {Sheth}(2002)}]{Cooray2002}
{Cooray}, A. \& {Sheth}, R. 2002, \physrep, 372, 1

\bibitem[{{Cora} {et~al.}(2018){Cora}, {Vega-Mart{\'\i}nez}, {Hough}, {Ruiz}, {Orsi}, {Mu{\~n}oz Arancibia}, {Gargiulo}, {Collacchioni}, {Padilla}, {Gottl{\"o}ber}, \& {Yepes}}]{cora_sag_2018}
{Cora}, S.~A., {Vega-Mart{\'\i}nez}, C.~A., {Hough}, T., {et~al.} 2018, \mnras, 479, 2

\bibitem[{{Dubinski} {et~al.}(1993){Dubinski}, {da Costa}, {Goldwirth}, {Lecar}, \& {Piran}}]{Dubinski1993}
{Dubinski}, J., {da Costa}, L.~N., {Goldwirth}, D.~S., {Lecar}, M., \& {Piran}, T. 1993, \apj, 410, 458

\bibitem[{Gottlober {et~al.}(2003)Gottlober, Lokas, Klypin, \& Hoffman}]{Gottlober_2003}
Gottlober, S., Lokas, E.~L., Klypin, A., \& Hoffman, Y. 2003, Monthly Notices of the Royal Astronomical Society, 344, 715–724

\bibitem[{{Hoyle} {et~al.}(2005){Hoyle}, {Rojas}, {Vogeley}, \& {Brinkmann}}]{Hoyle2005}
{Hoyle}, F., {Rojas}, R.~R., {Vogeley}, M.~S., \& {Brinkmann}, J. 2005, \apj, 620, 618

\bibitem[{{Hoyle} {et~al.}(2012){Hoyle}, {Vogeley}, \& {Pan}}]{Hoyle2012}
{Hoyle}, F., {Vogeley}, M.~S., \& {Pan}, D. 2012, \mnras, 426, 3041

\bibitem[{Hunter(2007)}]{Hunter2007}
Hunter, J.~D. 2007, Computing in Science \& Engineering, 9, 90

\bibitem[{Jian {et~al.}(2022)Jian, Lin, Hsieh, Lin, Umetsu, Lopez-Coba, Koyama, Hsu, Su, Chang, {et~al.}}]{jian2022}
Jian, H.-Y., Lin, L., Hsieh, B.-C., {et~al.} 2022, The Astrophysical Journal, 926, 115

\bibitem[{Jing {et~al.}(1998)Jing, Mo, \& B{\"o}rner}]{Jing1998}
Jing, Y., Mo, H., \& B{\"o}rner, G. 1998, \apj, 494, 1

\bibitem[{{Klypin} {et~al.}(2016){Klypin}, {Yepes}, {Gottl{\"o}ber}, {Prada}, \& {He{\ss}}}]{klypin_multidark_2016}
{Klypin}, A., {Yepes}, G., {Gottl{\"o}ber}, S., {Prada}, F., \& {He{\ss}}, S. 2016, \mnras, 457, 4340

\bibitem[{{Knebe} {et~al.}(2018){Knebe}, {Stoppacher}, {Prada}, {Behrens}, {Benson}, {Cora}, {Croton}, {Padilla}, {Ruiz}, {Sinha}, {Stevens}, {Vega-Mart{\'\i}nez}, {Behroozi}, {Gonzalez-Perez}, {Gottl{\"o}ber}, {Klypin}, {Yepes}, {Enke}, {Libeskind}, {Riebe}, \& {Steinmetz}}]{knebe_multidark_2018}
{Knebe}, A., {Stoppacher}, D., {Prada}, F., {et~al.} 2018, \mnras, 474, 5206

\bibitem[{Ma \& Fry(2000)}]{Ma2000}
Ma, C.-P. \& Fry, J.~N. 2000, \apj, 543, 503

\bibitem[{{Montero-Dorta} \& {Rodriguez}(2024)}]{Montero-Dorta_2024}
{Montero-Dorta}, A.~D. \& {Rodriguez}, F. 2024, \mnras, 531, 290

\bibitem[{{Padilla} {et~al.}(2005){Padilla}, {Ceccarelli}, \& {Lambas}}]{padilla_void_2005}
{Padilla}, N.~D., {Ceccarelli}, L., \& {Lambas}, D.~G. 2005, \mnras, 363, 977

\bibitem[{{Patiri} {et~al.}(2006){Patiri}, {Prada}, {Holtzman}, {Klypin}, \& {Betancort-Rijo}}]{Patiri2006}
{Patiri}, S.~G., {Prada}, F., {Holtzman}, J., {Klypin}, A., \& {Betancort-Rijo}, J. 2006, \mnras, 372, 1710

\bibitem[{Peacock \& Smith(2000)}]{Peacock2000}
Peacock, J. \& Smith, R. 2000, \mnras, 318, 1144

\bibitem[{{Perez} {et~al.}(2024){Perez}, {Pereyra}, {Coldwell}, {Rodriguez}, {Alfaro}, \& {Ruiz}}]{Perez2024}
{Perez}, N.~R., {Pereyra}, L.~A., {Coldwell}, G., {et~al.} 2024, \mnras, 528, 3186

\bibitem[{{Planck Collaboration} {et~al.}(2016){Planck Collaboration}, {Ade}, {Aghanim}, {Arnaud}, {Ashdown}, {Aumont}, {Baccigalupi}, {Banday}, {Barreiro}, {Bartlett}, {Bartolo}, {Battaner}, {Battye}, {Benabed}, {Beno{\^\i}t}, {Benoit-L{\'e}vy}, {Bernard}, {Bersanelli}, {Bielewicz}, {Bock}, {Bonaldi}, {Bonavera}, {Bond}, {Borrill}, {Bouchet}, {Boulanger}, {Bucher}, {Burigana}, {Butler}, {Calabrese}, {Cardoso}, {Catalano}, {Challinor}, {Chamballu}, {Chary}, {Chiang}, {Chluba}, {Christensen}, {Church}, {Clements}, {Colombi}, {Colombo}, {Combet}, {Coulais}, {Crill}, {Curto}, {Cuttaia}, {Danese}, {Davies}, {Davis}, {de Bernardis}, {de Rosa}, {de Zotti}, {Delabrouille}, {D{\'e}sert}, {Di Valentino}, {Dickinson}, {Diego}, {Dolag}, {Dole}, {Donzelli}, {Dor{\'e}}, {Douspis}, {Ducout}, {Dunkley}, {Dupac}, {Efstathiou}, {Elsner}, {En{\ss}lin}, {Eriksen}, {Farhang}, {Fergusson}, {Finelli}, {Forni}, {Frailis}, {Fraisse}, {Franceschi}, {Frejsel}, {Galeotta}, {Galli}, {Ganga}, {Gauthier}, {Gerbino}, {Ghosh}, {Giard},
  {Giraud-H{\'e}raud}, {Giusarma}, {Gjerl{\o}w}, {Gonz{\'a}lez-Nuevo}, {G{\'o}rski}, {Gratton}, {Gregorio}, {Gruppuso}, {Gudmundsson}, {Hamann}, {Hansen}, {Hanson}, {Harrison}, {Helou}, {Henrot-Versill{\'e}}, {Hern{\'a}ndez-Monteagudo}, {Herranz}, {Hildebrand t}, {Hivon}, {Hobson}, {Holmes}, {Hornstrup}, {Hovest}, {Huang}, {Huffenberger}, {Hurier}, {Jaffe}, {Jaffe}, {Jones}, {Juvela}, {Keih{\"a}nen}, {Keskitalo}, {Kisner}, {Kneissl}, {Knoche}, {Knox}, {Kunz}, {Kurki-Suonio}, {Lagache}, {L{\"a}hteenm{\"a}ki}, {Lamarre}, {Lasenby}, {Lattanzi}, {Lawrence}, {Leahy}, {Leonardi}, {Lesgourgues}, {Levrier}, {Lewis}, {Liguori}, {Lilje}, {Linden-V{\o}rnle}, {L{\'o}pez-Caniego}, {Lubin}, {Mac{\'\i}as-P{\'e}rez}, {Maggio}, {Maino}, {Mandolesi}, {Mangilli}, {Marchini}, {Maris}, {Martin}, {Martinelli}, {Mart{\'\i}nez-Gonz{\'a}lez}, {Masi}, {Matarrese}, {McGehee}, {Meinhold}, {Melchiorri}, {Melin}, {Mendes}, {Mennella}, {Migliaccio}, {Millea}, {Mitra}, {Miville-Desch{\^e}nes}, {Moneti}, {Montier}, {Morgante}, {Mortlock},
  {Moss}, {Munshi}, {Murphy}, {Naselsky}, {Nati}, {Natoli}, {Netterfield}, {N{\o}rgaard-Nielsen}, {Noviello}, {Novikov}, {Novikov}, {Oxborrow}, {Paci}, {Pagano}, {Pajot}, {Paladini}, {Paoletti}, {Partridge}, {Pasian}, {Patanchon}, {Pearson}, {Perdereau}, {Perotto}, {Perrotta}, {Pettorino}, {Piacentini}, {Piat}, {Pierpaoli}, {Pietrobon}, {Plaszczynski}, {Pointecouteau}, {Polenta}, {Popa}, {Pratt}, {Pr{\'e}zeau}, {Prunet}, {Puget}, {Rachen}, {Reach}, {Rebolo}, {Reinecke}, {Remazeilles}, {Renault}, {Renzi}, {Ristorcelli}, {Rocha}, {Rosset}, {Rossetti}, {Roudier}, {Rouill{\'e} d'Orfeuil}, {Rowan-Robinson}, {Rubi{\~n}o-Mart{\'\i}n}, {Rusholme}, {Said}, {Salvatelli}, {Salvati}, {Sandri}, {Santos}, {Savelainen}, {Savini}, {Scott}, {Seiffert}, {Serra}, {Shellard}, {Spencer}, {Spinelli}, {Stolyarov}, {Stompor}, {Sudiwala}, {Sunyaev}, {Sutton}, {Suur-Uski}, {Sygnet}, {Tauber}, {Terenzi}, {Toffolatti}, {Tomasi}, {Tristram}, {Trombetti}, {Tucci}, {Tuovinen}, {T{\"u}rler}, {Umana}, {Valenziano}, {Valiviita}, {Van Tent},
  {Vielva}, {Villa}, {Wade}, {Wandelt}, {Wehus}, {White}, {White}, {Wilkinson}, {Yvon}, {Zacchei}, \& {Zonca}}]{planck_2016}
{Planck Collaboration}, {Ade}, P.~A.~R., {Aghanim}, N., {et~al.} 2016, \aap, 594, A13

\bibitem[{{Prada} {et~al.}(2012){Prada}, {Klypin}, {Cuesta}, {Betancort-Rijo}, \& {Primack}}]{prada_mdpl_2012}
{Prada}, F., {Klypin}, A.~A., {Cuesta}, A.~J., {Betancort-Rijo}, J.~E., \& {Primack}, J. 2012, \mnras, 423, 3018

\bibitem[{{Ricciardelli} {et~al.}(2014){Ricciardelli}, {Cava}, {Varela}, \& {Quilis}}]{Ricciardelli_2014}
{Ricciardelli}, E., {Cava}, A., {Varela}, J., \& {Quilis}, V. 2014, \mnras, 445, 4045

\bibitem[{{Ricciardelli} {et~al.}(2017){Ricciardelli}, {Cava}, {Varela}, \& {Tamone}}]{Ricciardelli_2017}
{Ricciardelli}, E., {Cava}, A., {Varela}, J., \& {Tamone}, A. 2017, \apjl, 846, L4

\bibitem[{{Riebe} {et~al.}(2013){Riebe}, {Partl}, {Enke}, {Forero-Romero}, {Gottl{\"o}ber}, {Klypin}, {Lemson}, {Prada}, {Primack}, {Steinmetz}, \& {Turchaninov}}]{riebe_multidark_2013}
{Riebe}, K., {Partl}, A.~M., {Enke}, H., {et~al.} 2013, Astronomische Nachrichten, 334, 691

\bibitem[{{Rodriguez} \& {Merch{\'a}n}(2020)}]{RodriguezMerchan_2020}
{Rodriguez}, F. \& {Merch{\'a}n}, M. 2020, \aap, 636, A61

\bibitem[{{Rodriguez} {et~al.}(2015){Rodriguez}, {Merch{\'a}n}, \& {Sgr{\'o}}}]{Rodriguez2015}
{Rodriguez}, F., {Merch{\'a}n}, M., \& {Sgr{\'o}}, M.~A. 2015, \aap, 580, A86

\bibitem[{{Rodr{\'\i}guez-Medrano} {et~al.}(2023){Rodr{\'\i}guez-Medrano}, {Paz}, {Stasyszyn}, {Rodr{\'\i}guez}, {Ruiz}, \& {Merch{\'a}n}}]{Rodriguez-Medrano2023}
{Rodr{\'\i}guez-Medrano}, A.~M., {Paz}, D.~J., {Stasyszyn}, F.~A., {et~al.} 2023, \mnras, 521, 916

\bibitem[{{Rojas} {et~al.}(2004){Rojas}, {Vogeley}, {Hoyle}, \& {Brinkmann}}]{Rojas2004}
{Rojas}, R.~R., {Vogeley}, M.~S., {Hoyle}, F., \& {Brinkmann}, J. 2004, \apj, 617, 50

\bibitem[{{Rojas} {et~al.}(2005){Rojas}, {Vogeley}, {Hoyle}, \& {Brinkmann}}]{Rojas2005}
{Rojas}, R.~R., {Vogeley}, M.~S., {Hoyle}, F., \& {Brinkmann}, J. 2005, \apj, 624, 571

\bibitem[{Rosas-Guevara {et~al.}(2022)Rosas-Guevara, Tissera, Lagos, Paillas, \& Padilla}]{Rosas_Guevara_2022}
Rosas-Guevara, Y., Tissera, P., Lagos, C. d.~P., Paillas, E., \& Padilla, N. 2022, Monthly Notices of the Royal Astronomical Society, 517, 712–731

\bibitem[{{Ruiz} {et~al.}(2019){Ruiz}, {Alfaro}, \& {Garcia Lambas}}]{ruiz_into_2019}
{Ruiz}, A.~N., {Alfaro}, I.~G., \& {Garcia Lambas}, D. 2019, \mnras, 483, 4070

\bibitem[{{Ruiz} {et~al.}(2015){Ruiz}, {Paz}, {Lares}, {Luparello}, {Ceccarelli}, \& {Lambas}}]{ruiz_void_2015}
{Ruiz}, A.~N., {Paz}, D.~J., {Lares}, M., {et~al.} 2015, \mnras, 448, 1471

\bibitem[{Scoccimarro {et~al.}(2001)Scoccimarro, Sheth, Hui, \& Jain}]{Scoccimarro2001}
Scoccimarro, R., Sheth, R.~K., Hui, L., \& Jain, B. 2001, \apj, 546, 20

\bibitem[{Seljak(2000)}]{Seljak2000}
Seljak, U. 2000, \mnras, 318, 203

\bibitem[{{Shi} \& {Sheth}(2018)}]{shi2018}
{Shi}, J. \& {Sheth}, R.~K. 2018, \mnras, 473, 2486

\bibitem[{White \& Rees(1978)}]{White1978}
White, S.~D. \& Rees, M.~J. 1978, \mnras, 183, 341

\bibitem[{{Yang} {et~al.}(2007){Yang}, {Mo}, {van den Bosch}, {Pasquali}, {Li}, \& {Barden}}]{Yang2007}
{Yang}, X., {Mo}, H.~J., {van den Bosch}, F.~C., {et~al.} 2007, \apj, 671, 153

\bibitem[{Zehavi {et~al.}(2018)Zehavi, Contreras, Padilla, Smith, Baugh, \& Norberg}]{Zehavi2018}
Zehavi, I., Contreras, S., Padilla, N., {et~al.} 2018, \apj, 853, 84

\bibitem[{{Zheng} {et~al.}(2005){Zheng}, {Berlind}, {Weinberg}, {Benson}, {Baugh}, {Cole}, {Dav{\'e}}, {Frenk}, {Katz}, \& {Lacey}}]{Zheng2005}
{Zheng}, Z., {Berlind}, A.~A., {Weinberg}, D.~H., {et~al.} 2005, \apj, 633, 791

\end{thebibliography}

\end{document}